\documentclass[reprint,twocolumn,showpacs,aps,pra,amsmath,amssymb,superscriptaddress,floatfix,notitlepage]{revtex4-1}
\usepackage{graphicx} 
\usepackage{tikz}
\usepackage{placeins}

\newcommand{\ket}[1]{\vert #1 \rangle}
\newcommand{\bra}[1]{\langle #1 \vert}
\newcommand{\op}[2]{\vert#1\rangle\langle#2\vert}

\newcommand{\abs}[1]{\vert #1 \vert}

\newcommand{\tr}[2]{\text{tr}_#2\left\{#1\right\}}

\begin{document}
\title{Detecting non-Markovianity from continuous monitoring
}

\author{Kimmo Luoma}
\email{ktluom@utu.fi}
\affiliation{Turku Centre for Quantum Physics, Department of Physics and 
Astronomy, University of Turku, FI-20014, Turun Yliopisto, Finland}
\author{Pinja Haikka}
\affiliation{Department of Physics and Astronomy, 
Aarhus University, Ny Munkegade 120, DK-8000 Aarhus C, Denmark}
\author{Jyrki Piilo}
\affiliation{Turku Centre for Quantum Physics, Department of Physics and 
Astronomy, University of Turku, FI-20014, Turun Yliopisto, Finland}




\date{\today}

\begin{abstract}
We study how non-Markovianity of an open two-level system can be 
detected when continuously monitoring a part of its bosonic 
environment. Considering a physical scenario of an atom in a lossy 
cavity, we demonstrate that the properties of the time-dependent 
flux of the photons from the cavity allows the detection of memory effects in the atomic dynamics, without requiring 
state nor process tomography.
This framework overlaps with effective 
descriptions for the memory part of the environment using pseudomode methods. Our central results show how the 
Markovian measurement record on the environment of an enlarged open 
system allows to draw conclusions on the non-Markovianity of the 
original system of interest.  
\end{abstract}

\pacs{03.65.Yz, 42.50.Lc}
\maketitle
\section{Introduction}

During the last few years, several definitions and quantifiers for 
non-Markovianity of open quantum system dynamics have been 
introduced~\cite{PhysRevLett.101.150402,PhysRevLett.103.210401,
PhysRevLett.105.050403,fisherinfo,PhysRevA.86.044101,geomchar,2013arXiv1301.2585B,
PhysRevLett.112.120404,2014arXiv1408.7062B}. 
They are based on a number of different 
approaches, ranging from concepts of information 
flow~\cite{PhysRevLett.103.210401},
non-divisibility~\cite{PhysRevLett.105.050403}  
and  Fisher information~\cite{fisherinfo} to 
quantum mutual information~\cite{PhysRevA.86.044101}, 
accessible volume of physical 
states~\cite{geomchar}, channel capacity~\cite{2013arXiv1301.2585B}, 
and k-divisibility~\cite{PhysRevLett.112.120404}. Typically, the 
experimental detection of memory effects is difficult and requires either 
process~\cite{PhysRevLett.105.050403} or 
state tomography~\cite{PhysRevLett.103.210401,Piilo2011}. 
Despite recent progresses, developing simple schemes to 
experimentally detect non-Markovianity in open 
system dynamics remains a challenge. In this work we demonstrate how one can detect memory effects in the dynamics of 
a two-level system by 
continuously monitoring ~\cite{BRE02,carmichael2007statistical} a part of its 
environment and keeping the record of the arrival times of the photons.

The evolution of a Markovian open quantum
system is governed by the
celebrated Gorini-Kossakowski-Sudarshan-Lindblad master
equation~\cite{GKS76,Lindblad76}. This equation can be unravelled with 
different types of Markovian quantum trajectories 
\cite{PhysRevLett.68.580,PhysRevA.46.4363,PhysRevLett.82.1801,qsd2}, and
one can assign a certain physical reality to these trajectories 
in a sense that each trajectory represents the state of the open
quantum system conditioned on a particular measurement 
record of some environmental observable \cite{PhysRevA.47.642}.
The existence of such a measurement scheme interpretation for
unravelling of non-Markovian dynamics, instead, has been subject to much debate 
\cite{PhysRevA.58.1699,PhysRevA.61.062106,PhysRevA.66.012108,
PhysRevLett.100.180402,PhysRevLett.100.080401,PhysRevLett.101.140401}. 
Here we do not make claims about the existence (or non-existence) of the 
measurement scheme interpretation of conditional pure state evolutions 
for non-Markovian systems. However, we think that it is important to develop further 
legitimate schemes of measuring the environment
of an open system, whether it is Markovian or non-Markovian. 
In this work we show that a measurement record can, in fact, contain 
useful information regarding the detection of non-Markovianity 
without disturbing the open system dynamics.
Specifically, we study
an analytically solvable system where the environment
can be split into a memory and a non-memory part~\cite{PhysRevA.80.012104}. If
we then monitor the non-memory part of the environment,
this does not disturb the underlying non-Markovian
dynamics. Similar ideas have been studied
recently in Refs. \cite{PhysRevA.75.022103,diosi2012,PhysRevA.85.040101,
PhysRevA.88.012124,PhysRevLett.112.113601,2014arXiv1404.5280M}.

The outline of this paper is the following.  Section~\ref{sec:physical-model}
introduces our physical model, Sec.~\ref{sec:results} formulates the 
main result, and 
Sec.~\ref{sec:summary} concludes the Brief Report. 

\section{Physical model}\label{sec:physical-model}

Our physical system is a two-level atom interacting with a bosonic 
zero-temperature environment.
The Hamiltonian for this system, with $\hbar=1$, is 
$H=\omega_A\sigma_z +\sum_k\omega_k a_k^\dagger a_k+
\sum_kg_k(\sigma_-a_k^\dagger +\sigma_+a_k)$,
where $\omega_A$ is the transition frequency of atom, $\sigma_z$ is the 
Pauli $z$-matrix,
$a_k$ and $a_k^\dagger$ are the bosonic annihilation and creation operators
for the environmental modes, $g_k$ is the coupling between the atom
and the $k$-th field mode, and $\sigma_-$ and $\sigma_+$ are the 
lowering and raising operators for the atom.
In the continuum limit the sum over the environmental modes can be converted 
into an integral $\sum_k \mapsto \int d\omega_k \rho_k$, where 
$\rho_k$ is the density of states of the modes. 

We focus on a particular physical realization of this model, namely  
an atom inside a lossy optical cavity, so that the
coupling between the atom and the modes $\rho_kg_k^2$ is described 
by a Lorentzian spectral density
$\rho_kg_k^2 \equiv J(\omega_k)=
\frac{1}{2\pi}\frac{V^2\Gamma}{(\omega_k-\omega_c)^2+(\Gamma/2)^2}$.
Here, $\omega_c$ is the cavity resonance frequency, and $\Gamma$ and $V$ describe
the width and the strength of the spectral coupling, respectively.
If the modes are initially in the vacuum state, it is possible to
obtain an exact analytical time-local master equation for the 
two level atom only: $\dot{\rho}(t)=L_t\rho(t)$. The analytical form
of the superoperator $L_t$ is well known, and in certain cases it can 
describe non-Markovian atomic dynamics 
\cite{BRE02,PhysRevA.81.062124}.

This system can be mapped to a {\it Markovian}
system $\rho_{\rm{AP}}$ with the pseudomode method \cite{PhysRevA.55.2290},
where the two-level atom interacts only with a single
cavity mode, the pseudomode, which then leaks to a 
Markovian environment with a constant spectral coupling. 
The state of the two-level atom is obtained by tracing out
the pseudomode 
 $\rho=\tr{\rho_{\rm{AP}}}{P}$
 and it obeys exactly
the same dynamics as in the original picture. 
The Markovian master equation for the combined atom and pseudomode 
system, in the interaction
picture with respect to 
$H_A+H_{\rm{P}} = \omega_A\sigma_z+\omega_{\rm{P}} a_{\rm{P}}^\dagger a_{\rm{P}}$, is 
\begin{align}\label{eq:Master}
  \dot{\rho}_{\rm{AP}}(t)=&- i[H_0,\rho_{\rm{AP}}]\\
  &-
  \frac{\Gamma}{2}(a_{\rm{P}}^\dagger a_{\rm{P}} \rho_{\rm{AP}}+\rho_{\rm{AP}} 
a_{\rm{P}}^\dagger a_{\rm{P}} -2a_{\rm{P}}\rho_{\rm{AP}} a_{\rm{P}}^\dagger),\nonumber
\end{align}
where $H_0=V[e^{i\delta t}\sigma_-a_{\rm{P}}^\dagger+e^{-i\delta t}\sigma_+a_{\rm{P}}]$,
$a_{\rm{P}}$ and $a_{\rm{P}}^\dagger$ are the annihilation and creation operators of the 
pseudomode, $V$ and $\Gamma$ are the spectral coupling and the width of the 
original system, $\delta = \omega_{\rm{P}}-\omega_A$, and 
$H_A$ and $H_{\rm{P}}$ are the free Hamiltonians of the atom and the pseudomode.


\section{Results}\label{sec:results}

To solve the Markovian pseudomode master equation (\ref{eq:Master}) 
we introduce the combined (unnormalised) 
state $\ket{\tilde\Psi(t)}_{\rm{AP}} = c_0\ket{0}_A\ket{0}_{\rm{P}}+
c(t)\ket{1}_A\ket{0}_{\rm{P}}+b(t)\ket{0}_A\ket{1}_{\rm{P}}$ and find the 
following set of ordinary differential equations
\begin{align}
  \dot{c}(t) =& -i V e^{-i\delta t}b(t)\label{eq:A_amplitude},\\
  \dot{b}(t) =& -\frac{\Gamma}{2}b(t) -iVe^{i\delta t}c(t)\label{eq:PM_amplitude},
\end{align}
for the atom and pseudomode amplitudes, respectively, using a
technique presented in \cite{PhysRevA.55.2290}.
Coefficient $c_0$ is constant throughout the evolution and the 
analytical solutions of these equations are shown in Appendix \ref{sec:analyt-solut-c_1t}.

The pseudomode, described by operators 
$a_{\rm{P}}$ and $a_{\rm{P}}^\dagger$, can be interpreted as the memory part
of the environment, while the external modes, to which the pseudomode 
$a_{\rm{P}}$ decays, correspond to
the non-memory part of the environment \cite{PhysRevA.80.012104}. 
Our main result is that we can measure the non-memory part of 
the environment without disturbing the system of interest and extract
enough information from these measurements in order to decide whether the 
dynamics of the atomic system is Markovian or non-Markovian. 

We monitor continuously the photon flux $R(t)$ to the 
external modes.
From the Markovian Monte Carlo wave function (MCWF) 
method~\cite{PhysRevLett.68.580} we know that 
the probability for a quantum jump with jump operator $a_{\rm{P}}$ in 
Eq.~(\ref{eq:Master})
during small time interval $[t,t+\delta t)$ is 
$p(t)=\delta t \Gamma 
\bra{\varphi(t)}a_{\rm{P}}^\dagger a_{\rm{P}}\ket{\varphi(t)}=\delta t R(t)$, 
where $\ket{\varphi(t)}$ is a single trajectory obtained from the 
MCWF procedure.
This jump corresponds to the emission of a single 
photon to the external modes. It easy to show that the photon 
flux
from the pseudomode  is
$R(t) = \Gamma \bra{1}\tr{\rho_{\rm{AP}}(t)}{A}\ket{1}=
\Gamma\abs{b(t)}^2$, i.e.,
the product of the decay rate and the population of the pseudomode. 
Hamiltonian part of Eq.(\ref{eq:Master}) describes the coherent
excitation exchange between the atom and the pseudomode. The only 
way that the excitation can leak out to the 
external modes is through the pseudomode decay which may occur with
probability $p(t)$. After such event the state of the 
trajectory is $\ket{0}_A\ket{0}_{\rm{P}}$,
and stays there for the rest of the evolution.
This means that  we can continuously monitor the pseudomode decay without 
disturbing the non-Markovian dynamics of the atom system.

\begin{figure}
  \includegraphics[width=0.45\textwidth]{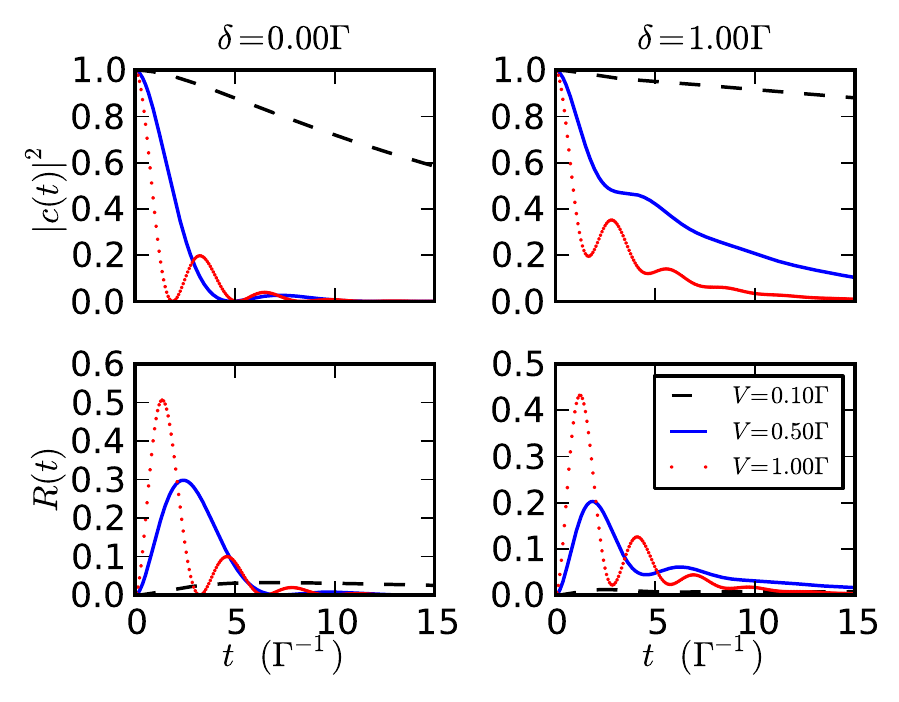}
  \caption{\label{fig:PopulationAndRate} (color online) 
    {\bf Top left}: Atom
    excited state population for different values of $V$ and for
    $\delta=0\Gamma$. {\bf Top right}: Atom
    excited state population for different values of $V$ and for
    $\delta=\Gamma$.
    {\bf Bottom left}: Photon flux for different values of
    $V$ and for $\delta=0\Gamma$.
    {\bf Bottom right}: Photon flux for different values of
    $V$ and for $\delta=\Gamma$. The legend here gives the values of 
    $V$ for all of the panels.}
\end{figure}

To detect the photon flux  experimentally one needs to be able to 
prepare 
the atom in an excited state $c(0)=1$ inside an empty cavity and 
then detect the times when a photon is emitted from 
the cavity. Then by repeating the procedure and time-binning 
the emission times one can construct $R(t)$ experimentally. We denote 
the total monitoring time with $T$. From now on initial conditions
are chosen as $c(0)=1$ and $b(0)=0$. With these fixed initial
conditions we can decide if the system is Markovian or non-Markovian 
for given values of the system parameters
(for details see Appendix \ref{sec:meas-non-mark}).

In Fig.~\ref{fig:PopulationAndRate}, we show the atomic excited state
population $|c(t)|^2$ and the photon flux $R(t)$ of the pseudomode 
for different values of coupling $V$ and detuning $\delta$
and for $T=14/\Gamma$.
When $\delta = 0$, the atom dynamics is non-Markovian for 
$V>\frac{\Gamma}{4}$. In the top left corner one can see oscillations 
of the excited
state populations, a signature of non-Markovian dynamics, 
for $V=\Gamma/2$ and $V=\Gamma$. In the bottom
left corner one can see oscillations emerging in $R(t)$ for
$V=\Gamma$ and very small amplitude oscillations also for $V=\Gamma/2$. 
For detuned case ($\delta=\Gamma$), right column of 
Fig. \ref{fig:PopulationAndRate}, atom is again non-Markovian 
for $V=\Gamma/2$ and we observe oscillations in 
the photon flux. The amplitude of $R(t)$ is now
smaller since the atom is depleting more slowly.
As one can see, whenever the atom dynamics is non-Markovian there 
are oscillations in the photon flux also. However,
oscillating photon flux $R(t)$ is not a sufficient
criteria to detect non-Markovianity correctly since 
$R(t)$ can oscillate when atom dynamics is Markovian 
if there is sufficient detuning between the atom and the
pseudomode transition frequencies 
(c.f. for example Fig.~\ref{fig:PopulationAndRate} blue line
on right panels).

 \begin{figure}
  \includegraphics[width=0.4\textwidth]{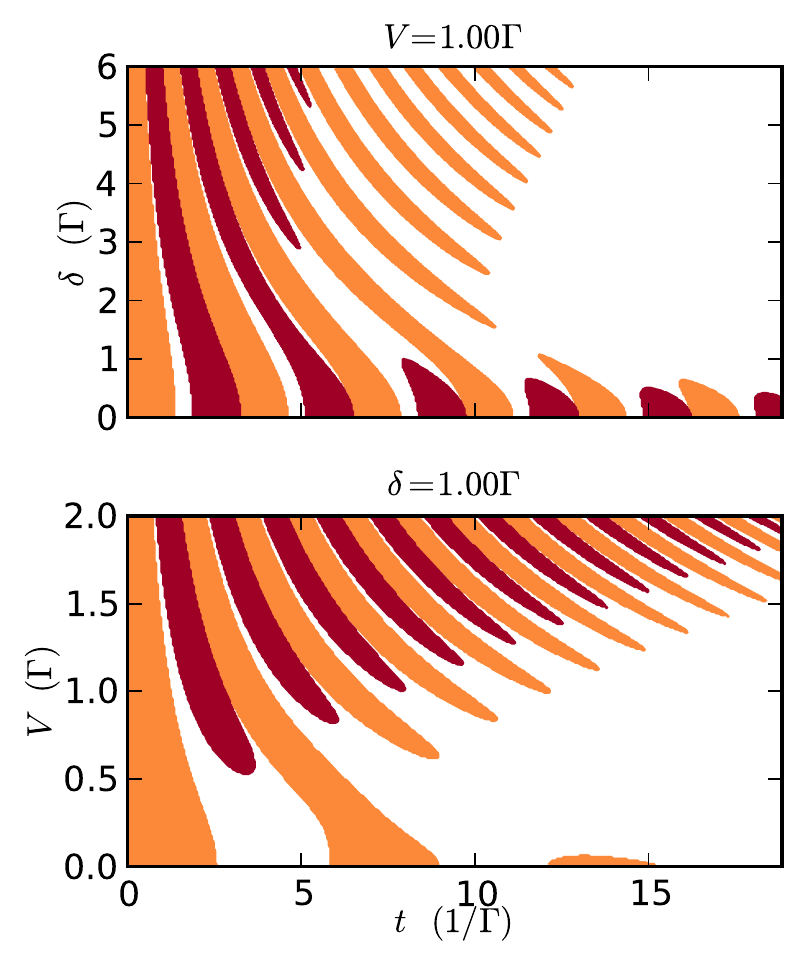}
  \caption{\label{fig:revivals}(color online) {\bf Top}: Regions in 
    $(t,\delta)$-plane where
    $C(t)=\partial_t \abs{c(t)}^2>0$ (dark red, black) and 
    $B(t)=\partial_t\Gamma\abs{b(t)}^2>0$ (orange, gray)
    for $V=\Gamma$.
    {\bf Bottom}: Regions in $(t,V)$-plane for $\delta=\Gamma$ where
    $C(t)>0$ or $B(t)>0$. Every revival of atomic population
    ($C(t)>0$) is followed by increase in the photon flux ($B(t)>0$).
    There are also areas where $B(t)>0$ without any previous
    atomic revivals. For all parameters photon flux increases initially.
  }
\end{figure}

In Fig.~\ref{fig:revivals} we plot the behavior of 
$C(t)=\partial_t \abs{c(t)}^2$ and $B(t)=\partial_t\Gamma\abs{b(t)}^2$ as 
a function of time. In the upper panel we have
fixed $V=\Gamma$. The dark red (black) 
areas in the $(t,\delta)$-plane correspond to $C(t)>0$, that is, 
the atom population 
is reviving. The orange (gray) areas, on the other hand, correspond 
to $B(t)>0$, that is an increasing emission rate.
In the lower panel we have repeated the calculation with a 
detuning of $\delta=\Gamma$. 

We observe that $B(t)$ is always initially positive, and that if 
there is a region where $C(t)>0$, this is always followed by a region 
with $B(t)>0$. This means that the atomic population revival is always 
followed by an increase of the photon flux. However, there are also 
regions where $B(t)>0$ without
preceding areas with $C(t)>0$ (lower panel with $V\approx \Gamma/2$), 
which means that the oscillating behavior of the
photon flux itself is not enough to decide whether the dynamics of the 
two-level atom is Markovian or non-Markovian.

 \begin{figure}
  \includegraphics[width=0.4\textwidth]{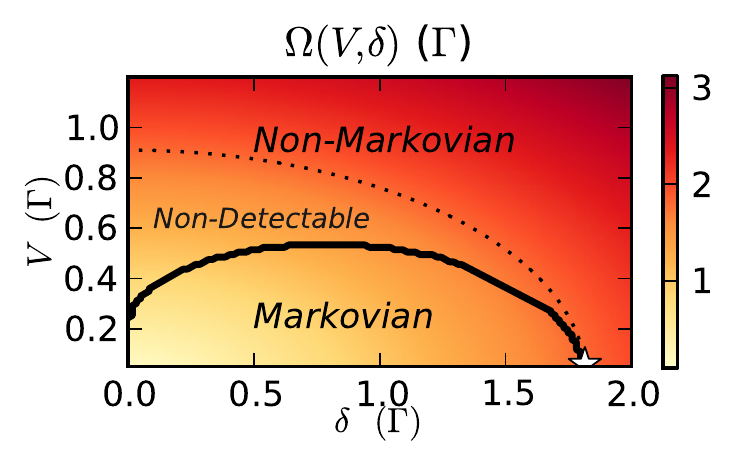}
  \caption{\label{fig:MaximumFreq} (color online) Oscillation frequencies
    $\Omega$ as a function of  $V$ and $\delta$
    near the transition from Markovian to non-Markovian dynamics. Below 
    boundary black solid boundary curve dynamics is Markovian.
    Above Markovian boundary the dynamics of open quantum system is 
    non-Markovian. 
    Maximal Markovian frequency  $\Omega_M\approx 1.8\Gamma$ is 
    marked with a white star. With dotted black we plot the contour
    where, $\Omega(V,\delta)=\Omega_M$. Non-Markovianity 
    can not be detected in the region between solid and dotted
    curves. In this figure $V\in[0.05\Gamma,1.2\Gamma]$.}
\end{figure}

\begin{figure}
  \includegraphics[width=0.4\textwidth]{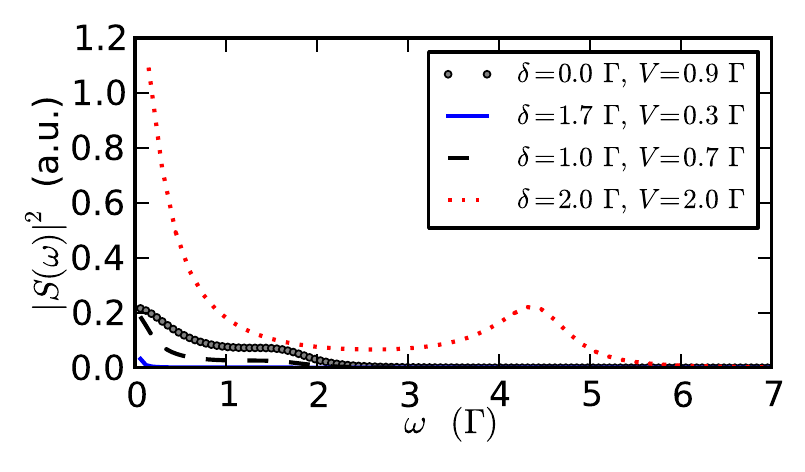}
  \caption{\label{fig:spectrum} (color online) Spectrum of the 
    photon flux.
    $\Omega_M\approx1.8\Gamma$ is the largest Markovian frequency. 
    In this figure
    we show evidence that for $(\delta,V)$ pair 
    $(2\Gamma,2\Gamma)$  (dotted red line) non-Markovianity is 
    detected because of a 
    pronounced peak in the flux occurring at $\omega \approx 4.47\Gamma$.
    Pair $(0.0\Gamma,0.9\Gamma)$ (dotted gray line) and $(\Gamma,0.7\Gamma)$  are just 
    below non-detectability border and they show some structure near 
    $\omega \approx \Omega_M$. $(1.7\Gamma,0.3\Gamma)$ (solid blue line)
    is also near detectability border and in the Markovian 
    region but there is no significant contribution to the 
    spectrum because amplitude of $R(t)$ is 
    very small, but there are oscillations.}
\end{figure}

Still, these findings suggest that non-Markovianity can be 
detected from the power spectrum $\abs{S(\omega)}^2$, which
contains information about the oscillation frequencies. 
We define the spectrum $S(\omega)$ as the Fourier transform of $r(t)=R(t)-\bar{R}_T$, 
where $\bar{R}_T=\frac{1}{T}\int_0^T R(s) \text{d}s$ is the time-average
of the photon flux and $T$ is the total time of observation.
The Wiener-Khinchin theorem states that the
power spectrum is related to the auto-correlation function
$\kappa(t)=\int_{-\infty}^\infty \text{d}\tau\,r(\tau)r(\tau+t)$ 
through the inverse Fourier transform (for more details,
see Appendix  \ref{sec:spectrum-somega}). 
It turns out that, when effect of damping is negligible, then 
there are periodic oscillations in the photon flux $R(t)$.   
The frequency of these 
oscillations is
\begin{align}
\Omega\equiv\Omega(V,\delta)=\sqrt{4V^2+\delta^2},
\end{align}  
Numerical evidence shows that the effect of damping $\Gamma$ is 
negligible to the oscillation frequency in 
the non-Markovian regime for short times. In the Markovian regime, 
the oscillation 
is damped after couple of cycles, has a 
small amplitude and is not periodic, hence it does not contribute to  
the power spectrum. In other words, when atom and pseudomode are 
coupled strongly enough, relative to the damping, there is coherent 
oscillation that can be observed in the spectrum and this occurs  
in the non-Markovian parameter regime.
Therefore using $\Omega$ and 
calculating maximum frequency $\Omega_M$ in the Markovian parameter
region, we obtain a threshold frequency that tells us that if there 
is coherent oscillations in the atom pseudomode dynamics with 
$\Omega>\Omega_M$, the atom dynamics is non-Markovian.

In Fig.~\ref{fig:MaximumFreq}
we show $\Omega(V,\delta)$ as a density plot with 
a boundary curve (solid black) between the
Markovian and non-Markovian regions in parameter space and 
with a contour curve $\Omega(V,\delta)=\Omega_M$ (dashed black).
We have also denoted the 
maximal Markovian frequency $\Omega_M\approx 1.8\Gamma$ with a
white star. Between Markovian boundary curve and contour 
curve corresponding to $\Omega_M$ there is region of ambiguity where
non-Markovianity could not be reliably detected. This occurs because 
Markovian dynamics could, in principle, contribute to the power spectrum
with these frequencies.

In Fig.~\ref{fig:spectrum} we study the behavior 
of the power spectrum $\abs{S(\omega)}^2$. We have calculated the 
spectrum in all three regions of Fig.~\ref{fig:MaximumFreq}.
Solid blue curve $(\delta,V)=(1.7\Gamma,0.3\Gamma)$ is in the 
Markovian region, dashed black  $(\Gamma,0.7\Gamma)$ and 
dotted gray curves $(0,0.9\Gamma)$ are 
in the non-detectable region and dotted red curve 
is in non-Markovian detectable region $(2\Gamma,2\Gamma)$.
In this figure we see that when the parameters 
are such that we are outside of the non-detection region we have 
a pronounced peak in the emission spectrum. We have also studied the 
behavior of the Markovian boundary near the 
non-detectability/non-Markovian boundary (solid blue curve) and it 
shows that choosing the threshold as we have done excludes false 
positive detections of non-Markovianity. What can also be deduced is 
that the pronounced behavior of the red dotted curve can be seen only
outside non-detection region when the open system exhibits 
non-Markovian dynamics. This figure confirms that threshold frequency 
$\Omega_M$ is compatible with non-Markovianity of the atom dynamics.



\section{Summary}\label{sec:summary}
To summarize, we have demonstrated that it is possible to detect 
the presence of memory effects for an open two-level system without 
doing a full state 
tomography -- provided that its non-Markovian dynamics is not 
perturbed by measurements on the environment. 
This was shown by mapping the original non-Markovian 
system to a larger Markovian one, where the memory and the 
non-memory parts
of the environment can be identified. For this purpose, 
we considered a two-level atom in a lossy cavity 
as a concrete example.

In the original system, non-Markovianity is caused by the
strong coupling between the atom and bosonic modes which
is manifested in the oscillations of the excited state population.
In the extended atom and single cavity mode system,
the atom and the cavity mode may exchange excitation coherently many 
times before the cavity mode is damped. This 
information is then encoded into the flux of the 
emitted photons from 
the cavity and its power spectrum.
As we showed, this information within the environment of the extended 
Markovian system allows us also to detect the presence of memory 
effects in the original open two-level system.

Our results are based on an interplay between the oscillations of the 
atomic excited state population and the flux of emitted 
photons from a cavity. It is expected that in the non-Markovian 
region the oscillations in the atomic excited state population also 
show up as oscillations in the photon flux from the cavity. 
However, in general the situation is more complicated since the time 
dependent photon flux  may also oscillate in the Markovian 
region when the atomic excited state population decreases 
monotonically. Subsequently, we have presented a spectrum analysis 
of the photon signal which allows to detect the presence of memory 
effects without directly measuring the state of the atom.  
In other words, there exists a threshold value in the frequency of 
the time dependent photon flux  oscillations which allows 
to study the Markovian and non-Markovian regions for the 
two-level atom dynamics.  
The experimental implementation of the scheme presented 
here is realistic with an ion trapped inside a cavity \cite{matthias}.

\begin{acknowledgements}
KL and JP  would like to thank Vilho, Yrj\"o, and Kalle V\"ais\"al\"a 
Foundation, and Jenny and Antti Wihuri Foundation for financial support.
PH acknowledges funding from the European Comission (ITN CCQED).
This research was undertaken on Finnish Grid Infrastructure (FGI) resources.

\end{acknowledgements}

\appendix

\section{Analytical solution for $c(t)$ and $b(t)$}\label{sec:analyt-solut-c_1t}
Solutions to Eqs. (\ref{eq:A_amplitude})-(\ref{eq:PM_amplitude}) 
with initial condition
$b(0)=0$ for cavity mode and generic initial condition $c(0)$ for
the excited state amplitude of the atom 
\begin{align}
  c(t) =& e^{-\frac{1}{4}t(\Gamma+2i\delta)}c(0)(\cosh(\frac{dt}{4})+\frac{\Gamma+2i\delta}{d}\sinh(\frac{dt}{4})),\\
  b(t) =& \frac{2i e^{-\frac{1}{4}t(d +\Gamma-2i\delta)} (e^{\frac{d t}{2}}-1) V c(0)}{d},\\
  d =& \sqrt{-16V^2+(\Gamma+2i\delta)^2}.
\end{align}
Solutions are obtained by Laplace transform method.

\section{Measure for non-Markovianity}\label{sec:meas-non-mark}
According to the reference~\cite{PhysRevLett.103.210401}, the 
dynamical map $\Phi_t$ describes non-Markovian evolution, if for 
some initial pair of states $\rho_{1,2}(0)$,  their trace distance 
$D\left(\rho_1(t),\rho_2(t) \right)= \frac{1}{2} \rm{tr} |\rho_1(t)-\rho_2(t)|$
increases temporally $\sigma(t)=\partial_tD(\rho_1(t),\rho_2(t))>0$. 
Measure for non-Markovianity is then obtained by 
\begin{align}
  \mathcal{N}=&\max_{\rho_{1,2}(0)}\int_{\sigma(t)>0}\sigma(t).
\end{align}
For our scheme, we want to obtain as large as possible photon flux
to the external modes of the extended Markovian system.
This is achieved  when initially the atom is maximally excited 
$\rho_{{\rm{AP}}}(0)=\op{1}{1}\otimes\op{0}{0}$ and $\rho_1(0)=\op{1}{1}$.
Maximizing pair for $\mathcal{N}$ must be orthogonal
\cite{PhysRevA.86.062108}, which leads in the considered purpose
$\rho_2(0)=\op{0}{0}$ and the trace distance is equal to the 
excited state population of $\rho_1(t)$
\begin{align}
  D(\rho_1(t),\rho_2(t))= \bra{1}\rho_1(t)\ket{1}=\abs{c(t)}^2.
\end{align}
For the current system, at least the divisibility 
measure~\cite{PhysRevLett.105.050403} and the above trace distance 
measure~\cite{PhysRevLett.103.210401}
are compatible since  there is only one decay channel and the dynamics 
becomes non-Markovian
when the corresponding decay rate becomes temporarily negative.

\section{Spectrum $S(\omega)$}\label{sec:spectrum-somega}
Let $r(t_i)=r_i\in \mathbb{R}$ be our experimental signal sampled at intervals 
$\delta t$ apart, $t_i=i\delta t,\, i = [0,N-1]$. 
Let sampling rate be $f_s$ and $\delta t= \frac{1}{f_s}$.
Discrete (only positive) frequencies are then 
$\frac{i f_s}{N}$, where $i\in [0,N/2-1]$ for $N$ even and
$i\in [0,(N-1)/2]$ for $N$ odd. 
Fourier transform of discrete signal is 
defined as  $S(\omega_k)=S_k = \sum_{m=0}^{N-1}r_me^{-2\pi i m k/N}$.
Inverse Fourier transform is defined as
$r_m=\frac{1}{N}\sum_{k=0}^{N-1}S_ke^{2\pi i m k/N}$. 
This leads to the following 
normalization $\sum_m \abs{r_m}^2 = \frac{1}{N} \sum_k \abs{S_k}^2$
(Parseval's theorem).
We have also the following relation
$\sum_{m}r_mr_{m+k}=\frac{1}{N}\sum_l\abs{S_l}^2e^{2\pi i kl/N}$
(Wiener-Kinchin theorem).

\bibliography{MeasuringNM}
\end{document}